\documentclass[aps, prd, showpacs, superscriptaddress, groupedaddress]{revtex4} 
\usepackage{graphicx}    
\usepackage{amssymb}
\usepackage{dcolumn}
\usepackage{color}
\usepackage{subfigure, rotating, bm, array}
\usepackage[pagebackref=false, colorlinks=true]{hyperref}
\hypersetup{
linkcolor=blue,     
citecolor=blue,     
urlcolor=blue}   
\usepackage{soul}
\usepackage{amsmath}
\begin{document}
\title{Globally visible singularity in an astrophysical setup}

\author{Karim Mosani}
\email{kmosani2014@gmail.com}
\affiliation{BITS Pilani K.K. Birla Goa Campus, Sancoale, Goa-403726, India}
\author{Dipanjan Dey}
\email{dipanjandey.icc@charusat.edu.in}
\affiliation{International Center for Cosmology, Charusat University, Anand 388421, Gujarat, India}
\author{Pankaj S. Joshi}
\email{psjprovost@charusat.ac.in}
\affiliation{International Center for Cosmology, Charusat University, Anand 388421, Gujarat, India}

\date{\today}

\begin{abstract}
The global visibility of a singularity as an end state of the gravitational collapse of a spherically symmetric pressureless cloud is investigated. We show the existence of a non-zero measured set of parameters: the total mass and the initial mean density of the collapsing cloud, giving rise to a physically strong globally visible singularity as the end state for a fixed velocity function. The existence of such a set indicates that such singularity is stable under small perturbation in the initial data causing its existence. This is true for marginally as well as non-marginally bound cases. The possibility of the presence of such suitable parameters in the astrophysical setup is then studied:
$1)$ The singularities' requirements at the center of the M87 galaxy and at the center of our galaxy (SgrA*) to be globally visible are discussed in terms of the initial size of the collapsing cloud forming them, presuming that such singularities are formed due to gravitational collapse.
$2)$ The requirement for the primordial singularities formed due to a collapsing configuration after getting detached from the background universe, at the time of matter-dominated era just after the time of matter-radiation equality, to be globally visible, is discussed.
$3)$ The scenario of the collapse of a neutron star after reaching a critical mass, which is achieved by accreting the supernova ejecta expelled by its binary companion core progenitor, is considered.
The primary aim of this paper is to show that globally visible singularities can form in astrophysical setups under appropriate circumstances.  
\end{abstract}
\maketitle
\section{Introduction}
Observations about compact objects such as the shadow of the M87 galactic center by the Event Horizon Telescope 
\cite{Akiyama}, 
earlier observations such as  Sagittarius A* (Sgr A*) at our galactic center 
\cite{Schodel, Ghez}, 
and similar compact objects at the center of other galaxies 
\cite{Kormendy} 
may hint that singularities do exist in our universe. General relativity predicts that a massive cloud under its gravitational influence collapses to form an infinitely dense singularity, curving the spacetime infinitely 
\cite{Hawking}. 
Escape of timelike or null geodesics may or may not be possible from such singularities. The weak cosmic censorship hypothesis, proposed by Penrose 
\cite{Penrose}, 
states that singularities cannot be visible to an asymptotic observer. However, it can now be shown that a collapsing dust cloud can give a globally visible singularity as its end state for certain mass profiles and velocity profiles
\cite{Deshingkar, Mosani}.

In addition to this, some theoretical predictions motivate us to consider the existence of a naked singularity. One of them is that shadows are not unique to black holes, but even naked singularities under certain circumstances can cast a shadow \cite{Shaikh}. 
Apart from this, the precession of the orbit of stars around the compact object at the galactic center can also put light on the causal structure of the central singularity. It has been found that  while the precession of particle orbit in the external Schwarzschild spacetime is in the direction of the particle motion, the naked singularity spacetimes like the Joshi-Malafarnia-Narayan (JMN) spacetime
\cite{JMN}
and the Janis-Newman-Winicour (JNW) spacetime
\cite{JNW}
can cause the orbiting particle to precess in the opposite direction of the particle's motion under certain conditions 
\cite{Bambhaniya, AJoshi}.
The S2 star orbiting around the center of our galaxy may also exhibit such behavior
\cite{Dey}.
Some observational signatures to look into, which help us in distinguishing from black holes and naked singularities, like the accretion disk property, the Einstein ring formed due to the gravitational lensing and the shadow, has been studied lately and can be found in 
\cite{Dey2, Dey3, Shaikh2, Bambi, Dey4}. 

The existence of a singularity may represent the incomplete nature of the classical theory of gravity 
\cite{Wheeler, HawkingCUP, Bergmann} 
and a quantum version of gravity possibly resolves it. However, if a singularity exists, it has its quantum aspects, which makes it more interesting to study. If such singularity allows null geodesic to escape from it so that it reaches an external observer without encountering trapped surfaces, then it could act as an astrophysical lab to test the quantum theory of gravity, which is still not understood
\cite{JoshiCUP}.

Mathematically, it is possible to achieve a globally naked singularity. However, whether suitable configurations required for its existence are possible astrophysically is our matter of concern, and we will discuss it in this paper. The paper is arranged as follows: Section II consists of the mathematical formalism of the gravitational collapse of a spherically symmetric dust cloud and the collapse criteria to end in a globally visible singularity. In section III, the astrophysically reasonable parameters, which can give a globally visible singularity, are discussed. The collapse of the cloud possibly forming the central singularity at the galactic center,  the collapse of the dark matter cloud in the primordial time, and the collapse of an accreting neutron star are investigated. We end the paper with the concluding remarks and open concerns in section IV. 

\section{Gravitational collapse formalism}

We consider the gravitational collapse of a spherically symmetric cloud of dust. The spacetime of such dust cloud is described by Lemaitre-Tolman-Bondi metric
\cite{Lemaitre, Tolman, Bondi}, 
which is given by
\begin{equation}
    ds^2=-c^2dt^2+\frac{R'}{1+f}dr^2+R^2d\Omega^2.
\end{equation}
Here $R=R(t,r)$ is the physical radius of the collapsing cloud and $f(r)$ is called the velocity function. The energy-momentum tensor in the comoving frame is given by
\begin{equation}
    T^{\mu\nu}=\rho(t,r) U^{\mu}U^{\nu}.
\end{equation}
where $\rho(t,r)$ is the energy density of the cloud  $U^{\mu}$ is the four velocity. Using the Einstein's field equations, we get the following relation:
\begin{equation}
    \frac{F'}{R^2R'}=8\pi \rho
\end{equation}
and
\begin{equation}\label{dotF}
    \dot F=0,
\end{equation}
where
\begin{equation}\label{friedmann}
    F=\frac{c^2}{G}\left(\frac{R \dot R^2}{c^2}-fR\right).
\end{equation}
The superscripts dot and prime denotes the partial derivative with respect to time and radial coordinate respectively. The term $F$ signifies the mass of the cloud inside a shell of radial coordinate $r$ at time $t$. It is called the Misner-Sharp mass function 
\cite{Misner}. 
In case of a collapsing dust, the Misner-Sharp mass function is a conserved quantity for a given radial coordinate and hence is a function of only $r$ as seen in Eq.(\ref{dotF}). 
Integrating Eq.(\ref{friedmann}), gives
\begin{equation}\label{t-tsr}
    t-t_s(r)=-\frac{R^{\frac{3}{2}}}{\sqrt{G}\sqrt{F}}\mathcal{G}\left(-\frac{c^2fR}{GF}\right),
\end{equation}
where $\mathcal{G}(y)$ is a positive, real and bound function given by
\begin{equation}\label{G}
\begin{split}
    &\mathcal{G}(y)=\left(\frac{\sin^{-1}\sqrt{y}}{y^{\frac{3}{2}}}-\frac{\sqrt{1-y}}{y}\right), \hspace{1cm} 0<y \leq 1, \\
    & \mathcal{G}(y)=\frac{2}{3}, \hspace{1cm} y=0.
    \end{split}
\end{equation}
Now, we can rescale the physical radius using the coordinate freedom such that
\begin{equation}
    R(0,r)=r.
\end{equation}
This along with Eq.(\ref{t-tsr}) gives
\begin{equation}\label{tsr}
    t_s(r)=\frac{r^{\frac{3}{2}}}{\sqrt{G}\sqrt{F}}\mathcal{G}\left(-\frac{c^2fr}{GF}\right),
\end{equation}
where $t_s(r)$ is the time taken by a shell of radial coordinate $r$ to collapse to a singularity. From Eq.(\ref{t-tsr}), Eq.(\ref{tsr}) and Eq.(\ref{G}), one can write $R$ explicitly as a function of $r$ and $t$ as
\begin{equation}\label{rnmb}
     R(t,r)_{\textrm{NMB}}=\frac{5GF}{2c^2f}\bigg (1-\bigg (1-\frac{4fc^2}{5FG}\bigg (r^{\frac{3}{2}}-\frac{3c^2f}{10GF}r^{\frac{5}{2}}-\frac{3}{2}\sqrt{G}\sqrt{F}t\bigg )^{\frac{2}{3}}\bigg )^{\frac{1}{2}}\bigg),
\end{equation}
for $f\neq 0$. The subscript ``NMB" stands for non-marginally bound collapse. Here, we have used the Taylor expansion representation of $\mathcal{G}(y)$ up to first order, which is given by
\begin{equation}\label{taylorexpansionofG}
    \mathcal{G}(y)=\frac{2}{3}+\frac{y}{5}+o(y^2), \hspace{1cm} 0<y<1.
\end{equation}
Eq.(\ref{rnmb}) is a good approximation only if the velocity function is such that $|f|<<\frac{G F}{c^2 r}$. Velocity functions with higher magnitude require higher orders to be considered in the Taylor expansion of $\mathcal{G}(y)$.
For the case of marginally bound (MB) collapse, $R$ is expressed as
\begin{equation}\label{rmb}
    R(t,r)_{\textrm{MB}}=\left(r^{\frac{3}{2}}-\frac{3}{2}\sqrt{G}\sqrt{F}t\right)^{\frac{2}{3}}.
\end{equation}
The condition for the apparent horizon curve can be obtained from the following equality:
\begin{equation}
    g^{\mu\nu}R,_{\mu}R,_{\nu}=0.
\end{equation}
This gives the following relation needed to be satisfied for the apparent horizon curve:
\begin{equation}
    \frac{\dot R^2}{c^2}=1+f.
\end{equation}
The above equation along with Eq.(\ref{friedmann}) gives
\begin{equation} 
    F=\frac{c^2}{G}R.
\end{equation}
The event horizon evolves like a null geodesic wavefront and meets the apparent horizon at the boundary of the collapsing cloud $r_c$. Hence, the event horizon is described by the solution of the differential equation:
\begin{equation}\label{ngde}
    \frac{dt}{dr}=\frac{1}{c}\frac{R'}{\sqrt{1+f}},
\end{equation}
satisfying the condition:
\begin{equation}\label{ngdeic}
    F(r_c)=\frac{c^2}{G}R(t,r_c).
\end{equation}
We consider the Misner-Sharp mass function to be of the form
\begin{equation}\label{strongmassfunction}
    F=F_0\left(\frac{r}{r_c}\right)^3+F_3\left(\frac{r}{r_c}\right)^6.
\end{equation}
According to Tipler 
\cite{Ellis, Tipler}, the volume element formed by three independent Jacobi fields along timelike geodesics terminating at a physically strong singularity should vanish. Using the sufficient criteria by Clarke and Krolak
\cite{Clarke}
for a singularity to be Tipler strong, one can conclude that in the case of a marginally bound collapse, the mass function as stated in Eq.(\ref{strongmassfunction}) maintains the strength of the singularity in the sense of Tipler
\cite{Joshi}. 
Additionally, for $F_3<0$, such singularity is locally naked as well. In the case of non-marginally bound collapse, the strength of the singularity and its local causal property depends on the velocity function in addition to the mass function
\cite{Mosani} 
(Please refer to the Appendix for further discussion on the local visibility of the singularity).    

For the density to decrease as we move away from the center and smoothly vanish at the boundary of the collapsing cloud ($\rho(0,r_c)=0$), $F$ has to be rewritten as 
\begin{equation}\label{msmf}
    F(r)=F_0\left(\frac{r}{r_c}\right)^3\left(1-\frac{1}{2}\left(\frac{r}{r_c}\right)^3\right).
\end{equation}
It should be noted that in the case of marginally bound pressureless collapsing cloud with density vanishing smoothly at its boundary, the above expression is the only possible mass function which corresponds to a Tipler strong singularity as its end state (See the Appendix).

Substituting Eq.(\ref{rnmb}) and Eq.(\ref{msmf}) in Eq.(\ref{ngdeic}), we get 
\begin{widetext}
\begin{equation}\label{tehrc}
    t_{\textrm{EH}_{\textrm{NMB}}}(r_c)=\frac{2\sqrt{2}}{3\sqrt{G}\sqrt{F_0}}r_C^{\frac{3}{2}}-\frac{2\sqrt{2}}{5}\frac{c^2f(r_c)}{\left(G F_0\right)^{\frac{3}{2}}}-\frac{G F_0}{3c^3}\left(1-\frac{f(r_c)}{5}\right)^{\frac{3}{2}}
\end{equation}
\end{widetext}
for $f\neq 0$. For $f=0$, substituting Eq.(\ref{rmb}) and Eq.(\ref{msmf}) in Eq.(\ref{ngdeic}), we get
\begin{equation}\label{tehrc2}
    t_{\textrm{EH}_{\textrm{MB}}}(r_c)=\frac{2\sqrt{2}}{3\sqrt{F_0}\sqrt{G}}\left(r_c^{\frac{3}{2}}-\left(\frac{GF_0}{2c^2}\right)^{\frac{3}{2}}\right).
\end{equation}
\begin{figure*}\label{fig1..}
\subfigure[]
{\includegraphics[scale=0.5]{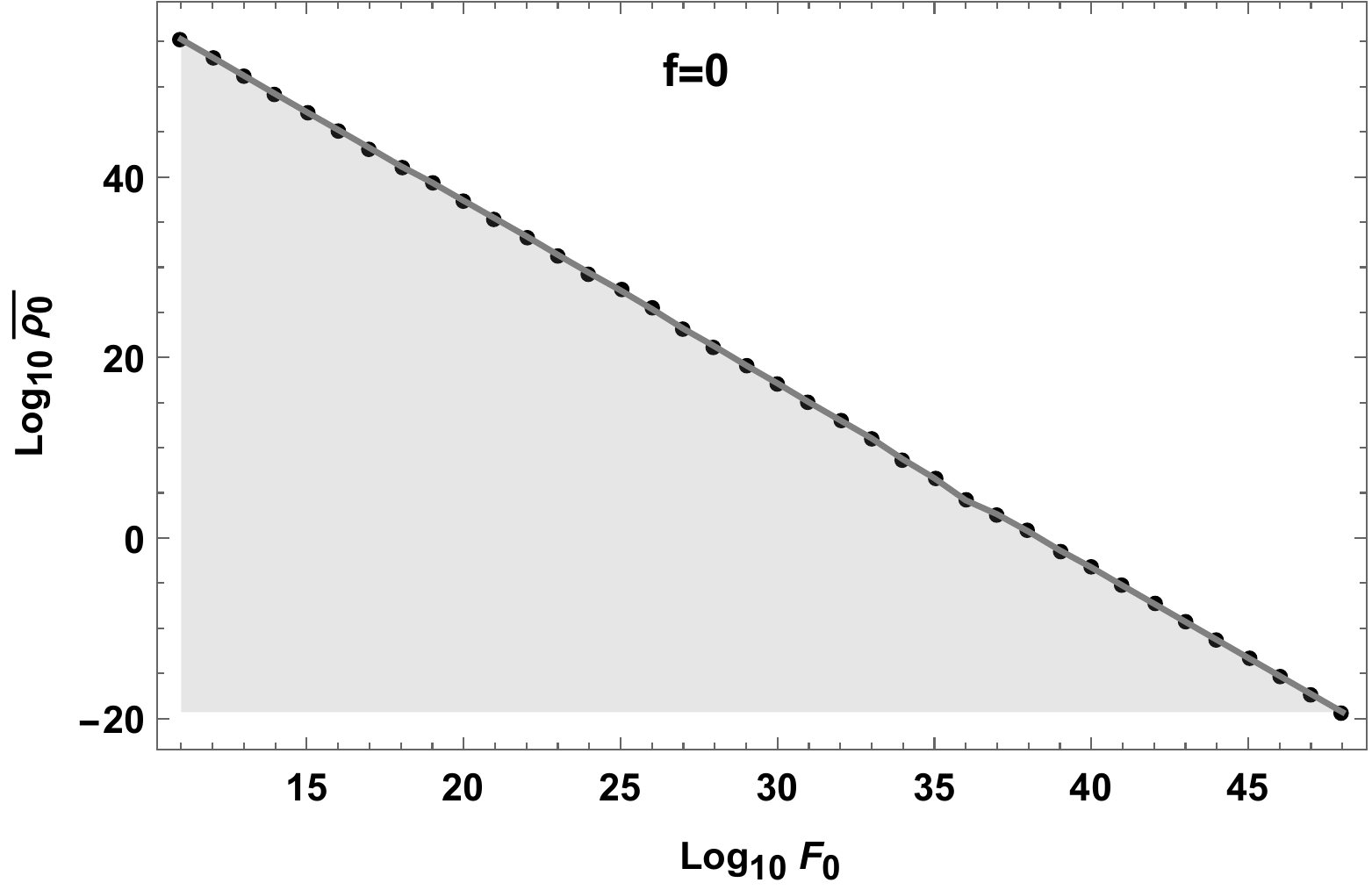}}
\hspace{0.2cm}
\subfigure[]
{\includegraphics[scale=0.5]{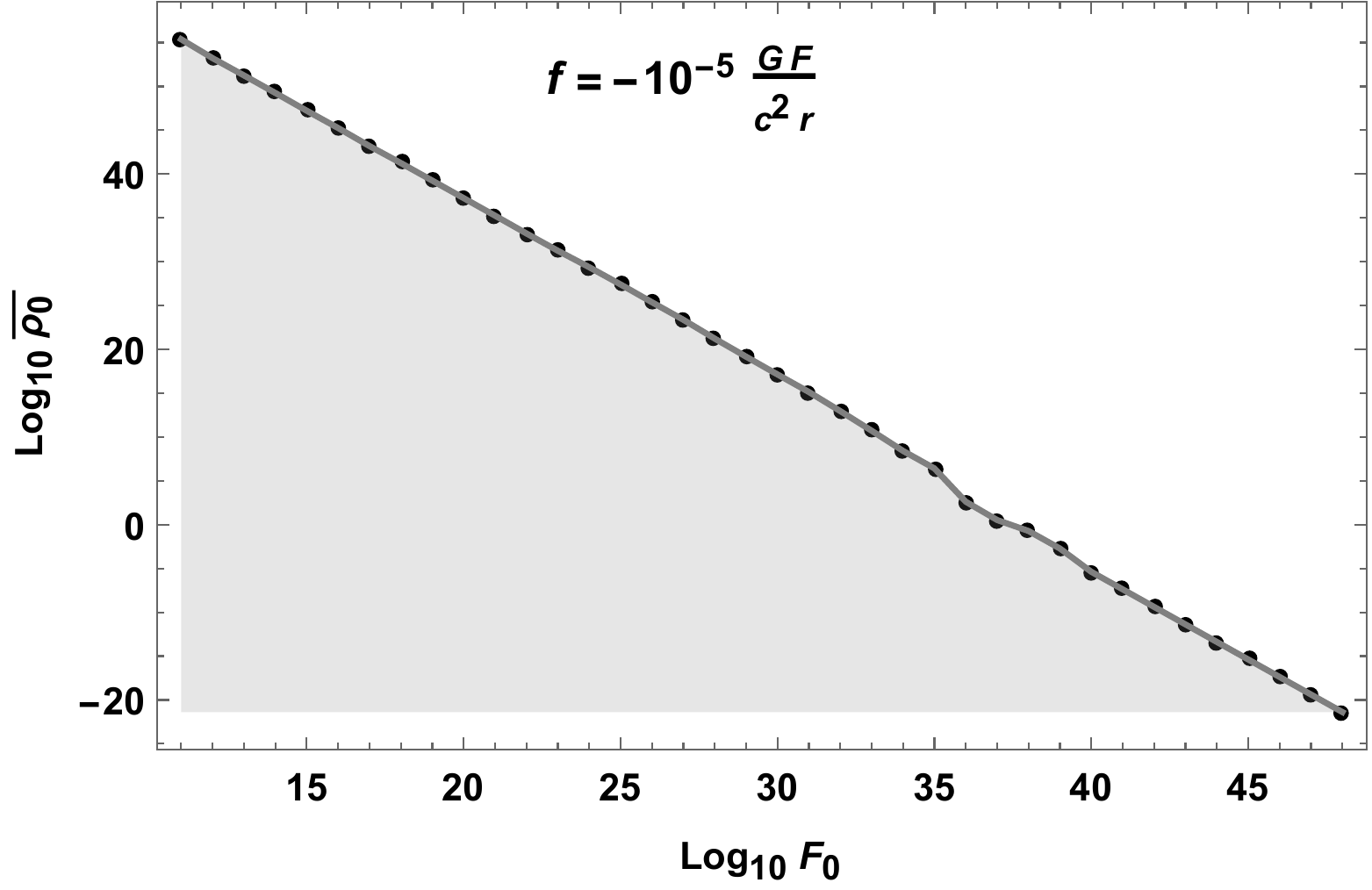}}
\caption{Non-zero measured set of initial data giving rise to a globally visible singularity (shaded region) and globally hidden singularity (unshaded region) as an end state of a marginally bound collapse (a) and a non-marginally bound collapse (b) of a pressureless inhomogeneous dust cloud having the mass function as in Eq.(\ref{msmf}). Here, $\Bar{\rho_0}$ and $\frac{F_0}{4}$ are the initial mean density and the total mass of the collapsing cloud having units $\textrm{kg}/\textrm{m}^3$ and kg respectively.  }
\end{figure*}
The event horizon curve $t_{EH}(r)$ is now the solution of the differential Eq.(\ref{ngde}) satisfying Eq.(\ref{tehrc}) in case of non-marginally bound collapse (Eq.(\ref{tehrc2}) in case of marginally bound case). At the time of formation of the central singularity, the apparent horizon (the boundary of all trapped surfaces) starts forming from the center. It evolves in the outward direction for an inhomogeneous mass function as in Eq.(\ref{msmf}) \cite{Mosani, Joshi2}. Hence, any null geodesic leaving the center after the formation of the singularity, due to the collapse of the central shell, gets trapped by trapped surfaces. Therefore, the event horizon forms either before or during the formation of the central singularity. If it forms at the center before the formation of this central singularity, any null geodesic escaping from the center after $t_{EH}(0)$, having a positive tangent at $r=0$, will fall back to the singularity, thereby making the singularity globally hidden. Hence, the necessary criteria for a singularity to be globally visible is as follows: 
\begin{equation}\label{ts0equalteh0}
    t_s(0)=t_{\textrm{EH}}(0). 
\end{equation}
We note here that the singularity has to be a nodal point apart from satisfying the above equality. This is to make sure that an entire family of null geodesic escapes from the point $(t_s(0),0)$ in the $(t,R)$ plane so that the central singularity remains visible to an asymptotic observer for an infinite time 
\cite{Joshi2}. 
These escaping null geodesics are solutions of the differential equation (\ref{ngde}) starting from the point $(t_s(0),0)$ in the $(t,R)$ plane, and reaching the boundary $r_c$ before the event horizon, for which it has to satisfy the following inequality:
\begin{equation}
    F(r_c)<\frac{c^2}{G}R(t,r_c).
\end{equation}. We now use Eq.(\ref{ts0equalteh0}) to plot Fig.(1) on the basis of which we study the astrophysical relevance of globally visible singularity in the following section.

\section{Astrophysical relevance}
It is evident that for a pressureless marginally bound collapsing cloud, with smoothly vanishing density at its boundary, ending up in a Tipler strong singularity, the global visibility or otherwise of the singularity depends on two parameters: the total mass $\frac{F_0}{4}$ and the initial size $r_c$ of the collapsing cloud at the beginning of its collapse (or equivalently $\frac{F_0}{4}$ and its initial mean density $\Bar{\rho_0}=\frac{3F_0}{8\pi r_{c}^{3}}$). This is because, in the case, $f=0$, the equality or otherwise of the singularity curve Eq.(\ref{tsr}) and the event horizon curve (which is a solution of the differential Eq.(\ref{ngde}) satisfying the condition Eq.(\ref{ngdeic})), which decides the global visibility of the singularity, depends on only $F_0$ and $r_c$. 
A similar situation may or may not occur in the case of a non-marginally bound collapse.  It is possible to show some examples of the mass profile, values of $F_0$, and $\Bar{\rho_0}$, which gives rise to the globally visible singularity as an end state. However, do such values arise in nature? Fig. (1) is a plot of $\textrm{log}_{10}F_0$ v/s $\textrm{log}_{10}\Bar{\rho_0}$ (for marginally bound case as well as non marginally bound case) in which the dotted line is the cut off for $\Bar{\rho_0}$ for a given $F_0$ having the property that any $\Bar{\rho_0}$ less than this cut off gives a globally visible singularity for a given $F_0$. In the case of non-marginally bound collapse (Fig(1.b)), the velocity function is taken such that mass and size remain the only variables that control the global causal property of the singularity. Also, it is small enough to avoid error due to the approximation of Taylor expansion of Eq.(\ref{taylorexpansionofG}) up to only the first order. Here, we study three types of astrophysical scenarios giving rise to a singularity and investigate if they can be observed by an asymptotic observer. 
\begin{figure*}\label{fig1}
\subfigure[]
{\includegraphics[scale=0.5]{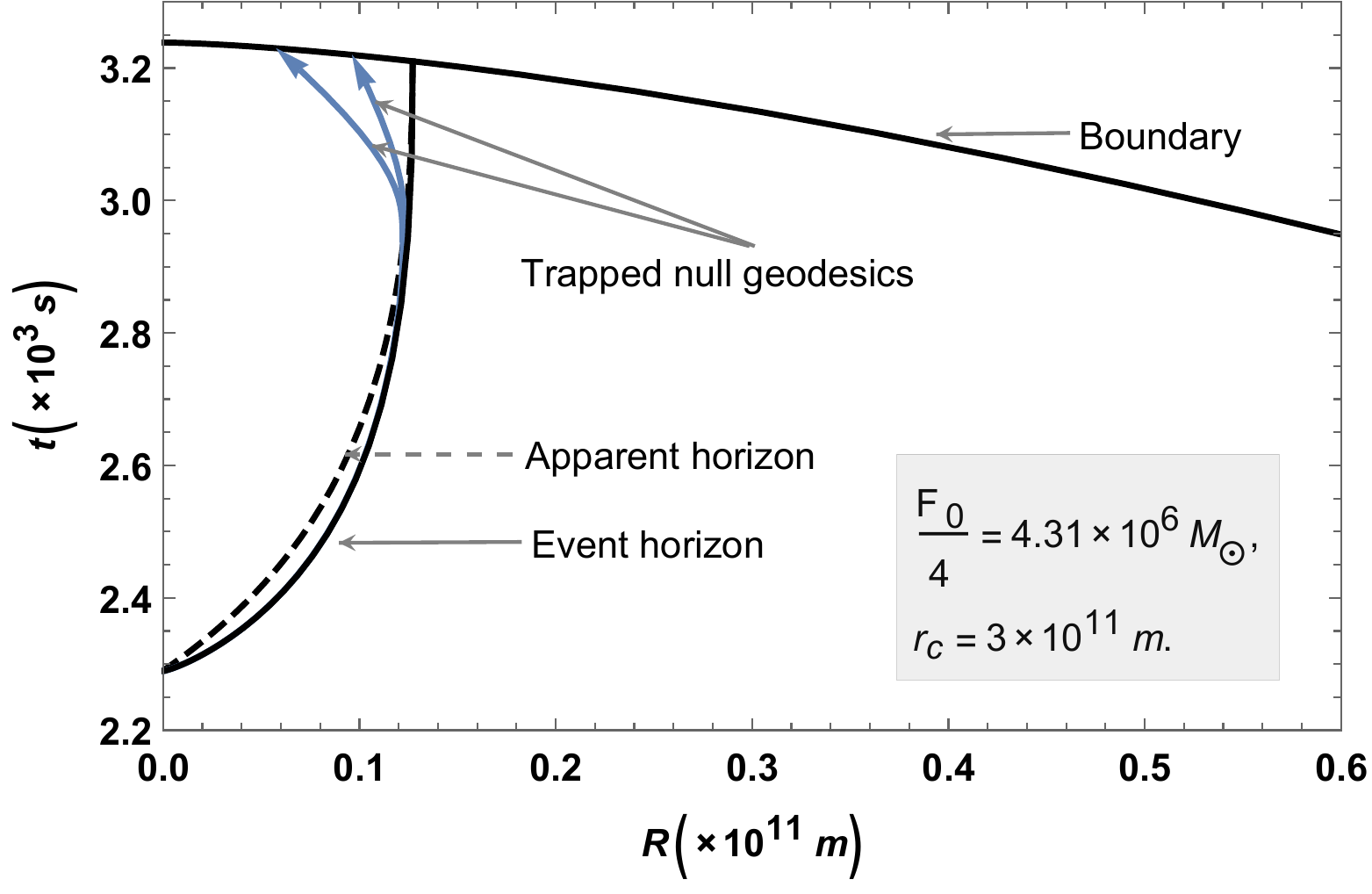}}
\hspace{0.2cm}
\subfigure[]
{\includegraphics[scale=0.5]{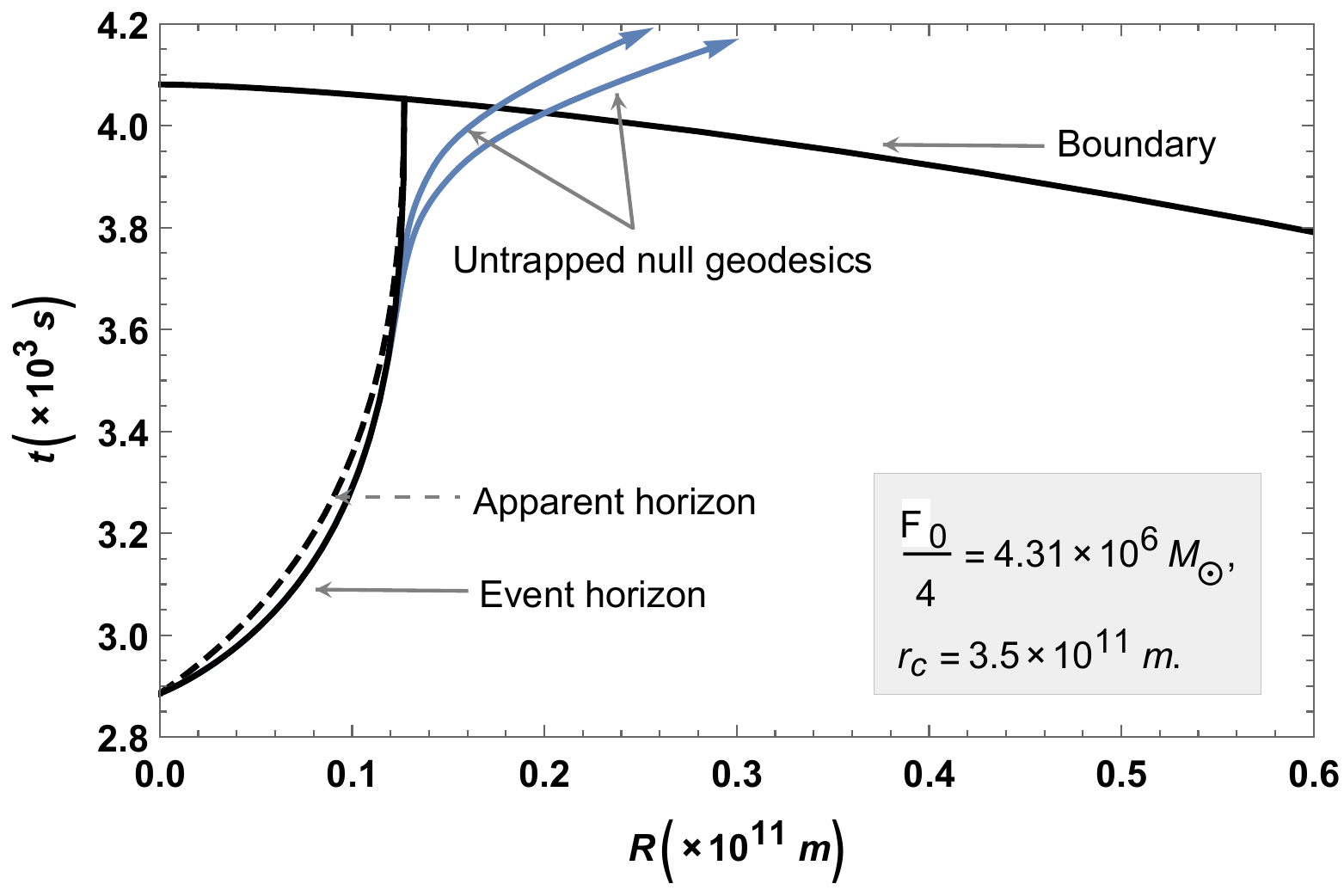}}
\caption{Causal structure of space-time singularity formed at the center of the milky way galaxy (the compact object SgrA*). The initial unknown radius of the collapsing core forming such singularity decides its end state. $r_c\leq 3.32\times 10^{11} \textrm{m}$ and $r_c \geq 3.33\times 10^{11} \textrm{m}$ ends up globally hidden and a globally visible singularity respectively. Collapsing core is assumed to be spherical, pressureless and marginally bound. Mass function is chosen such that the singularity is Tipler strong.}
\end{figure*}
\begin{figure*}\label{fig3}
\subfigure[]
{\includegraphics[scale=0.5]{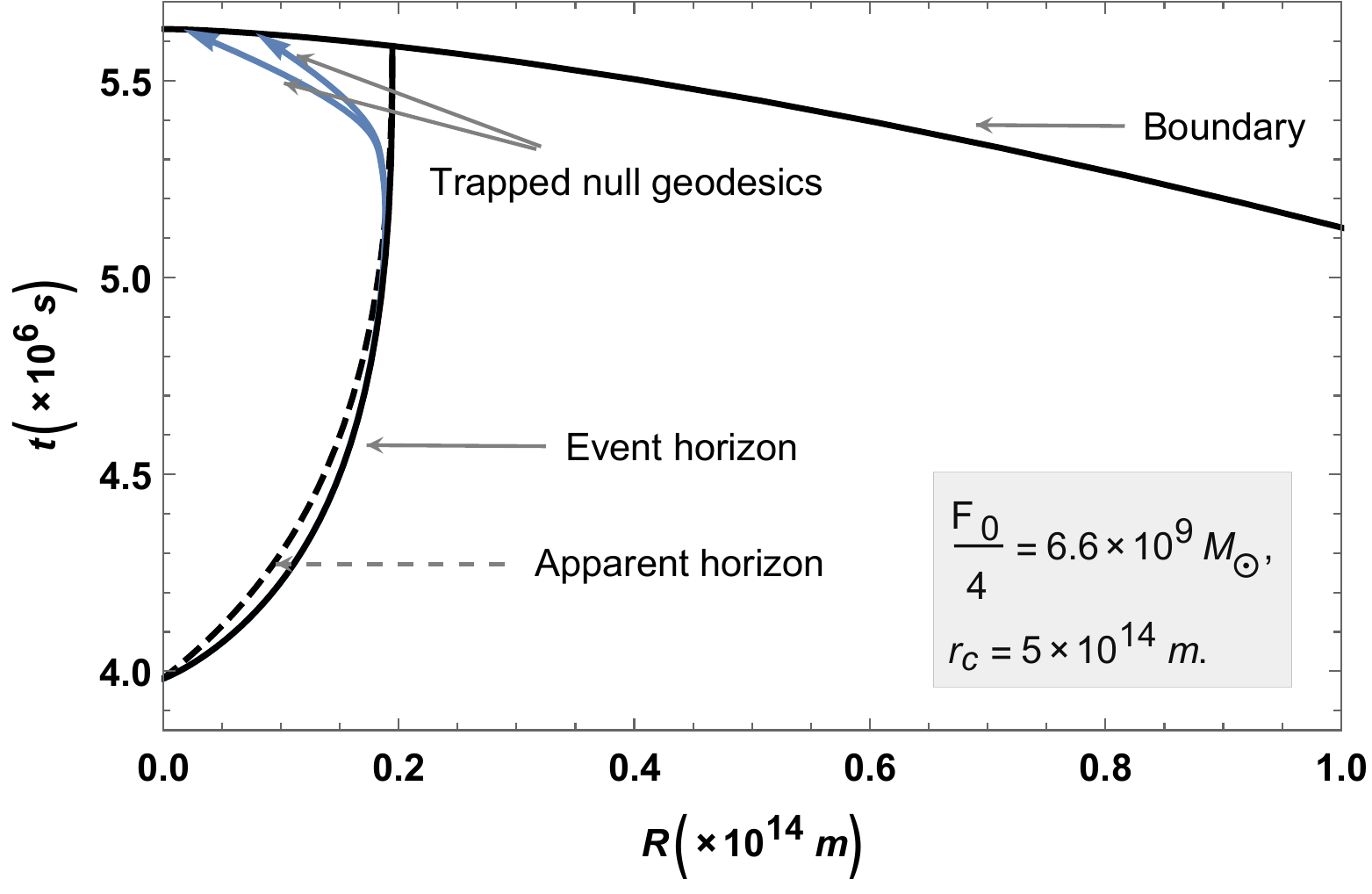}}
\hspace{0.2cm}
\subfigure[]
{\includegraphics[scale=0.5]{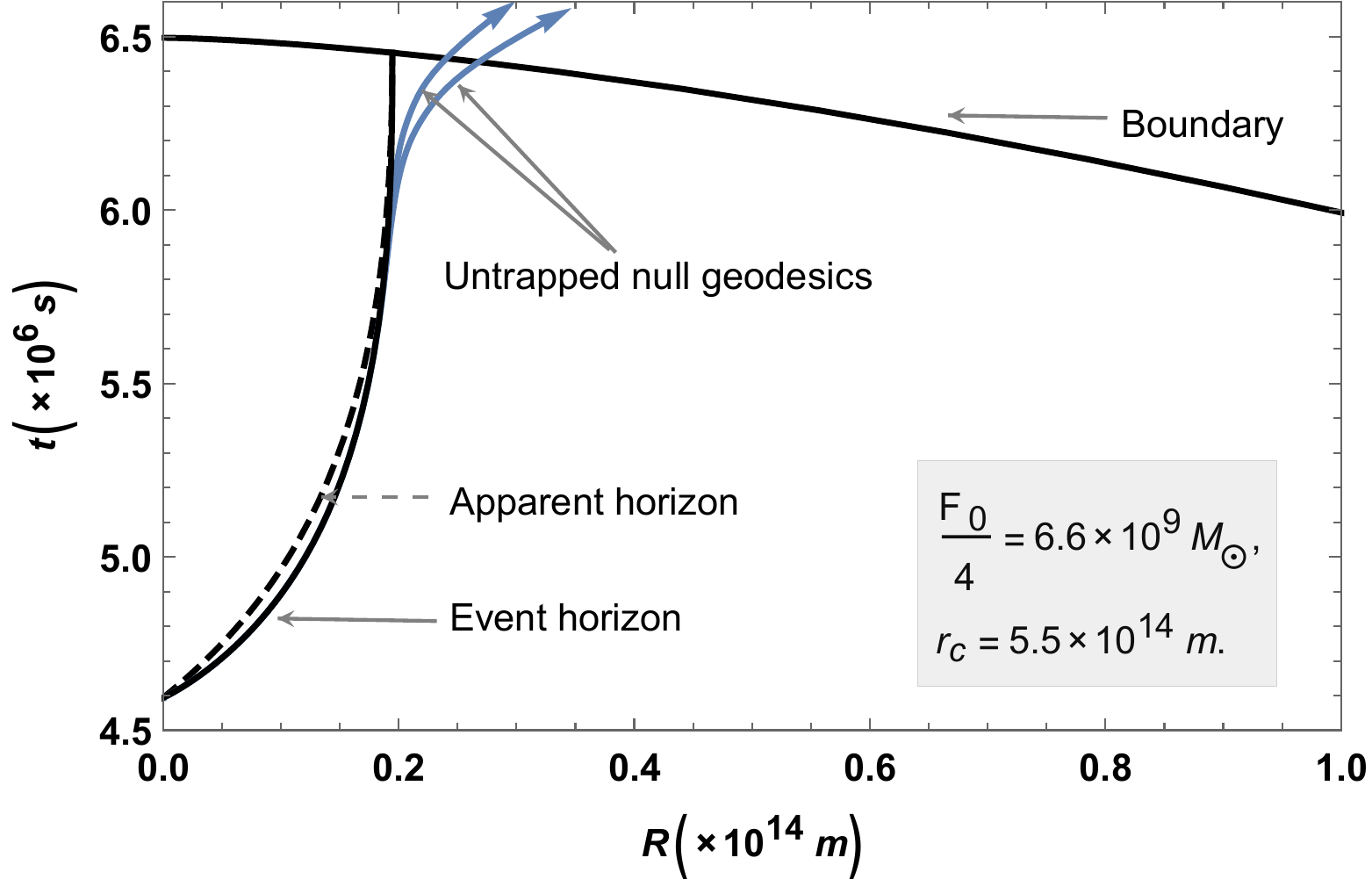}}
\caption{Causal structure of space-time singularity formed at the center of the M87 galaxy. The initial unknown radius of the collapsing core forming such singularity decides its end state. $r_c\leq 5.13\times 10^{14} \textrm{m}$ and $r_c \geq 5.14\times 10^{14} \textrm{m}$ ends up in a globally hidden and a globally visible singularity respectively. Collapsing core is assumed to be spherical, pressureless and marginally bound. Mass function is chosen such that the singularity is Tipler strong.}
\end{figure*}

\subsection{The galactic center}
We investigate the causal structure of the central singularity of SgrA*, which is a compact object at the center of our galaxy. The mass of this object as calculated by Gillessen \textit{et. al.} 
\cite{Gillessen}
is $4.31 \times 10^6 M_{\odot}$, and its Schwarzschild radius is $\sim 1.26 \times 10^{10}$ m. For the SgrA* singularity to be globally visible, the collapsing spherical cloud forming such singularity should have the initial radius greater than equal to $\sim 3.33\times 10^{11}$ m (in case of a marginally bound case), which translates to roughly at least $26.42$ times the size of its Schwarzschild radius. A lesser initial radius will make the singularity globally hidden. This corresponds to having the initial mean density less than $55.42 \textrm{kg}/\textrm{m}^3$. Fig.(2) is the visual representation of both the possible outcomes.

Similarly, the compact object at the center of the M87 galaxy has mass $6.6 \times 10^9 M_{\odot}$ as calculated by the Event Horizon Telescope collaboration
\cite{Akiyama2},
and a Schwarzschild radius $\sim 1.9 \times 10^{13}$ m. For the central singularity to be globally visible, the collapsing spherical cloud forming such singularity should have the initial radius greater than or equal to $\sim 5.14\times 10^{14}$ m (in case of marginally bound collapse), which roughly translates to at least $27.05$ times its Schwarzschild radius. A lesser initial radius will make the singularity globally hidden. This corresponds to having the initial mean density less than $2.31 \times 10^{-5} \textrm{kg}/\textrm{m}^3$. Fig.(3) is the visual representation of both the possible outcomes.

\subsection{Singularities having primordial origin}
Another astrophysically motivated entity that we consider here is the primordial singularities. Primordial fluctuations in the very early universe could give rise to such singularities
\cite{Zeldovich, Hawking2}. For a primordial singularity to form due to gravitational collapse, we investigate the collapse of matter in the era just after the matter-radiation equality, i.e., the start of the matter-dominated era. The reason for considering this time as the initiation of the collapse is that the temperature during this epoch drops down to $9000$ K, and the dark matter configuration can now collapse. We know that the epoch of matter-radiation equality occurred at redshift $z\sim 3000$
\cite{Frieman}. 
For a $\Lambda\textrm{CDM}$ model, the deceleration to acceleration transition of the universe is calculated to begin at $z\sim 0.72$
\cite{Farooq}. The relation between the density of the universe at these epochs depends on $z$ at both these epochs as follows:
\begin{equation}\label{rhoz}
   \frac{\rho_{\textrm{MRE}}}{\rho_{\textrm{DA}}}=\left (\frac{1+z_{\textrm{MRE}}}{1+z_{\textrm{DA}}}\right )^{3}. 
\end{equation}
This is because the era bounded below and above by the time of matter-radiation equality and the time of deceleration to acceleration transition, respectively, has matter as the dominant fluid, and the density of the universe is dependent on the linear equation of state parameter $\omega$ of the dominating fluid, as follows
\cite{Liddle}:
\begin{equation}
    \rho(\omega) \propto \left(1+z\right)^{3(1+\omega)}.
\end{equation}
Hence for $\omega=0$ (dust domination), Eq.(\ref{rhoz}) holds. If we consider the dark energy candidate driving the present accelerated expansion as the cosmological constant ($\omega=-1$), its density $\rho_{\Lambda}$ remains constant in time. Therefore, we can say that
\begin{equation}\label{rhodalambda}
    \rho_{\textrm{DA}}=\rho_{\Lambda}.
\end{equation}
 Using the above equality in Eq.(\ref{rhoz}), one can obtain the density of the universe at the time of matter-radiation equality as
 \begin{equation}
    \rho_{\textrm{MRE}}= 3.092 \times 10^{-17} \textrm{kg}/\textrm{m}^3.
 \end{equation}
Here, we have used the present Hubble parameter, $H_0=67.4 \textrm{km s}^{-1}\textrm{Mpc}^{-1}$, and the density parameter of $\Lambda$, $\Omega_{\Lambda}=0.686$, from the Planck 2018 results
\cite{Planck}
to get the value of $\rho_{\Lambda}= 5.82298 \times 10^{-27} \textrm{kg}/\textrm{m}^3$ using the relation $\rho_{\Lambda}=\frac{3H_0^2\Omega_{\Lambda}}{8\pi G}$
\cite{Liddle}.
The overdensity configuration, formed due to primordial fluctuation just after the epoch of matter-radiation-equality, detaches itself from the background universe and starts contracting when the density contrast $\delta \rho/ \rho$ reaches order one
\cite{Khlopov}. 
Hence, at the start of the collapse, one can assume the order of the initial mean density of the collapsing configuration, which we denote by $\rho_{\textrm{CONFIG}}$, same as the order of the density of the background universe, i.e. $10^{-17} \, \textrm{kg}/\textrm{m}^3$. Therefore, we have 
\begin{equation}
    \rho_{\textrm{CONFIG}}< \Bar{\rho_0} \left (\frac{F_0}{4} \right )
\end{equation}
for all configuration having $\frac{F_0}{4}< 10^{45} \,\textrm{kg}$ as is evident from Fig.(1) in both marginally bound as well as non-marginally bound case for very small velocity function. Hence, we can state that any configuration with $\frac{\delta \rho}{\rho}$ having order one, and having mass less than $10^{45}\, \textrm{kg}$ collapses to give a globally visible singularity if the collapse initiates just after the time $t_{\textrm{MRE}}$. It is worth mentioning that the overdense region collapses even if the density contrast is of the order greater than one, in which case, the result may vary accordingly. 

\subsection{Collapse of a matter accreting neutron star}
We now consider a close binary system scenario, one of which is a star undergoing supernova explosion and another is a neutron star, as investigated by Rueda and Ruffini \cite{Reuda}. We give a brief overview of this scenario.

The material expelled from the core progenitor is accreted by the companion neutron star and can get captured if it falls in the region at a distance less than
\begin{equation}
    R_{cap}=\frac{2GM_{NS}}{v_{rel}^{2}}
\end{equation} from the center of the neutron star. Here, $M_{NS}$ is the mass of the neutron star, and $v_{rel}$ is the velocity of the ejected particle relative to the neutron star orbital motion, and is given by 
\begin{equation}
    v_{rel}=\sqrt{v^{2}_{orb}+v^{2}_{ej}},
\end{equation} 
where 
\begin{equation}
v_{orb}=\sqrt{\frac{G(M_{prog}+M_{NS})}{a}}.
\end{equation}
Here, $v_{ej}$ is the ejecta velocity, which at the start of the supernova explosion has a value of the order of $10^7$ $m s^{-1}$, $v_{orb}$ is the orbit velocity of the neutron star, $M_{prog}$ is the mass of the progenitor, and $a$ is the binary separation. We neglect the neutron star magnetic field's effect by considering the magnetospheric radius $R_{m}$ comparatively smaller than $R_{cap}$. This is possible for a suitably high rate of accretion since 
\begin{equation}
    R_{m}=\frac{B^2R_{NS}^6}{(\dot M \sqrt{2G M_{NS}})^{\frac{2}{7}}},
\end{equation}
where $\dot M=\frac{dM}{dt}$ is the matter accretion rate of the supernova ejecta by the neutron star, and $B$, $M_{NS}$ and $R_{NS}$ are the magnetic field, mass and the radius of the neutron star respectively 
\cite{Toropina}. 
The mass accretion rate is given by 
\cite{Hoyle} 
\begin{equation}\label{accretionrate}
    \dot M= \epsilon \pi \rho_{ej} R^{2}_{cap}v_{rel}=\epsilon \pi \rho_{ej}\frac{(2GM_{NS})^2}{v^{3}_{rel}}.
\end{equation}
Here, $0<\epsilon<1$, and depends on the medium in which the accretion process takes place. $\rho_{ej}(t)=\frac{3M_{ej}(t)}{4 \pi r_{ej}^{3}(t)}$ is the density of the ejected material which decreases with time. The supernova ejecta radius can be assumed to expand as $r_{ej}=a t^b$, where $a$ and $b$ are constants \cite{Chevalier}. Eq.(\ref{accretionrate}) can now be integrated to get the magnitude of the mass accreted by the neutron star. After some time, provided the accretion rate is high,in order that the effect of the magnetic field is negligible, the mass of the neutron star reaches the critical mass and further collapses unhindered. Neutron star stable configuations have been studied taking into account weak, strong, electromagnetic and gravitational interaction in the framework of general relativity 
\cite{Belvedere}. 
Tab.(1) mentions the critical mass $M_{crit}$ and its corresponding radii for some parameterizations of neutron star models studied in 
\cite{Belvedere}. 
If we model the unhindered collapse of the neutron star after achieving the critical masses mentioned in Tab.1, by a spherical pressureless marginally bound cloud collapse, having the mass function of the form  of Eq.(\ref{msmf}), then each model (i.e. $NL3$ \cite{Lalazissis}, $NL-SH$\cite{Sharma}, $TM1$ \cite{Sugahara}, and $TM2$ \cite{Hirata}) ends up in a singularity which is hidden globally. This is because the event horizon forms before the singularity at $r=0$ in each case as seen using equations (\ref{tsr}), (\ref{ngde}) and (\ref{ngdeic}). However, the singularity is visible locally, as explained in the appendix.
\begin{table}[t]
\begin{tabular}{|l|c|c|c|c|c|}
\hline
Model  & $M_{crit}$ ($\times M_{\odot}$) &  $R$ ($km$) & $t_s(0)$ $(\times 10^{-5} s)$ & $t_{EH}(0)$ $(\times 10^{-5} s)$\\
\hline
\hline
    NL3     & $2.67$  & $12.33$ & $3.42937$  &  $0.83442$  \\
\hline
   NL-SH       & $2.68$   & $12.54$ & $3.51078$ & $0.91288$  \\
\hline
   TM1    & $2.58$  & $12.31$ & $3.48018$ & $0.98919$   \\
\hline 
   TM2     & $2.82$  &  $13.28$ & $3.72990$ & $1.00047$ \\
\hline
 \end{tabular}
\caption{Neutron star critical masses and the corresponding radii are obtained by fixing certain nuclear parameters, which includes the coupling constants and the meson masses \cite{Belvedere}. Each of these four models corresponds to one such fixed-parameter set. For each model, in the case of marginally bound unhindered gravitational collapse, the event horizon forms before the formation of the singularity at $r=0$. Hence the singularity thus formed is hidden globally.}
\label{tab1}
\end{table}
\begin{figure}\label{fig4}
\includegraphics[scale=0.5]{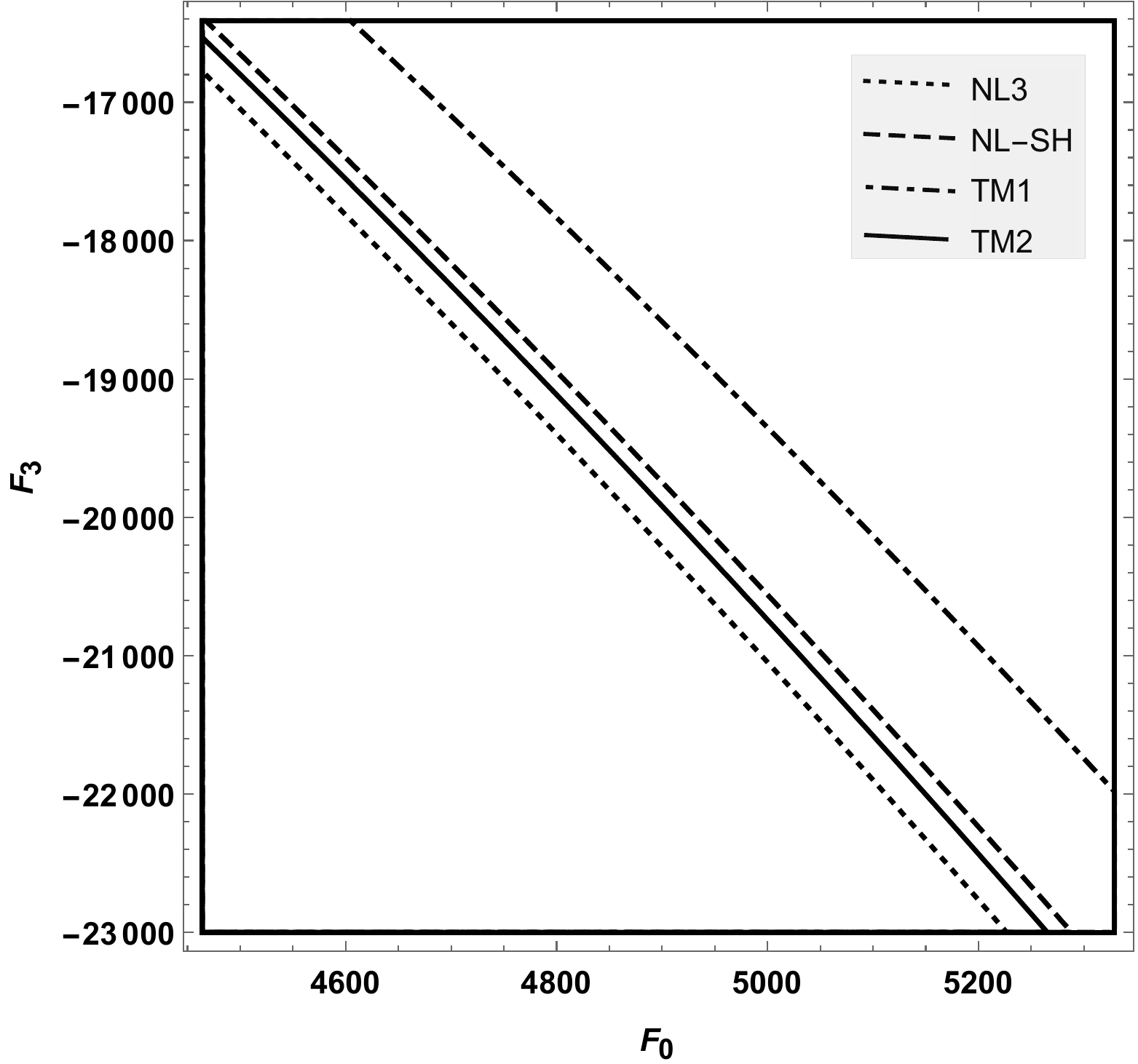}
\caption{Neutron star critical masses and the corresponding radii are obtained by fixing certain nuclear parameters which includes the coupling constants and the meson masses \cite{Belvedere}. Each of the four models: $NL3$, $NL-SH$, $TM1$, and $TM2$ corresponds to one such fixed parameter set. The Misner-Sharp mass function is chosen as $F(r,R)=F_0\left(\frac{r}{r_c}\right)^3+F_3\left(\frac{r}{r_c}\right)^6+F_R\left(\frac{R}{r_c}\right)^3$. We have considered a marginally bound collapse, i.e. $\mathcal{Y}=1$. The total initial mass, i.e. the critical mass of the collapsing cloud is given by $M_{crit}=(F_0+F_3+F_R)/2$. The magnitudes of $F_0$ and $F_3$ depicted in the plot are considerably smaller than that of $F_R$. Hence, $M_{crit}$ is approximately $F_R/2$. The region of parameter space $(F_0,F_3)$ below (above) the curve gives a globally visible (hidden) singularity at the end of the collapse of a neutron star after reaching the critical mass by accreting the supernova ejecta from its binary companion core progenitor.}
\end{figure}
One could argue that modeling the cloud formed due to matter accreting neutron star by zero pressure is not justified since neutron star has non-zero pressure, following a certain equation of state. To investigate the end state of the collapsing cloud formed after the neutron star accretes the supernova ejecta from its binary companion core progenitor, we rely on the work by Giambo \textit{et al.}
\cite{Giambo, Giambo2}. 
We give a brief overview of the formalism presented in these articles.

The general spacetime metric of a spherical collapsing cloud in the comoving coordinates is given by
\begin{equation}
    ds^2=-c^2\frac{\dot R^2}{\mathcal{H}}dt^2+\frac{R'^2}{\mathcal{Y}}dr^2+R^2d\Omega^2,
\end{equation}
where $R$, $\mathcal{Y}$ and $\mathcal{H}$ are functions of $t$ and $r$. The above metric can be rewritten in the transformed comoving area-radial coordinates $(r,R)$ as
\begin{equation}
     ds^2=-Adr^2-2BdRdr-\frac{1}{\mathcal{H}}dR^2+R^2d\Omega^2,
\end{equation}
where $A$ and $B$ are functions of $r$ and $R$. $\mathcal{H}$ can also be written as a function of $r$ and $R$ as
\begin{equation}
    \mathcal{H}(r,R)=\frac{F(r,R)}{R}+\mathcal{Y}(r,R)^2-1.
\end{equation}
The transformed metric is advantageous to obtain information about the global causal structure of the singularity.
A quantity $\Delta(r,R)$ is defined as
\begin{equation}
\Delta=B^2-\frac{A}{\mathcal{H}}= \frac{R'^2}{\mathcal{Y} \mathcal{H}}.
\end{equation}
Using the Einsteins' field equation, one can get an integral expression of $\sqrt{\Delta}$ as
\begin{equation}
    \sqrt{\Delta}=\int_R^r \frac{1}{\sqrt{\mathcal{Y}(r,R)}}\frac{\partial}{\partial r}\left(\frac{1}{\mathcal{H}(r,R)}\right)dR+\frac{1}{\sqrt{\mathcal{Y}(r,r)\mathcal{H}(r,r)}}.
\end{equation}
The formalism is restricted to only those cases where the energy density $\rho$ and matter density $\rho_m$ are related as 
\begin{equation}
    \rho=\omega \rho_m c^2,
\end{equation}
where 
\begin{equation}
    \omega=\frac{E(r)}{2}\left(\frac{F_r(r,R)}{\sqrt{\mathcal{Y}(r,R)}}+\frac{F_R(r,R)R'(r,R)}{\sqrt{\mathcal{Y}(r,R)}} \right).
\end{equation}
The energy density, the radial pressure, and the tangential pressure are respectively given by
\begin{equation}
\rho=\frac{c^2}{8\pi R^2}\left(\frac{F,_r}{R'}+F,_R\right),
\end{equation}
\begin{equation}\label{prgiambo}
p_r=-\frac{c^2 F,_R}{8\pi R^2},
\end{equation}
and
\begin{equation}
p_t=-\frac{c^2}{16\pi R R'}\left(F,_{rR}-\sqrt{\mathcal{Y}} F,_r\frac{\partial}{\partial R}\left(\sqrt{\mathcal{Y}}\right)+F,_{RR}R'\right).
\end{equation}
In a collapsing cloud, provided: $(1)$ the weak energy condition hold, $(2)$ the regularity conditions hold, $(3)$ the shell crossing singularity do not occur, and $(4)$ the shell focusing singularity form in a finite comoving time, the singularity formed as an end state of the gravitational collapse is globally visible if $n=3$ and $\frac{\zeta}{\alpha}>\frac{26+15\sqrt{3}}{2}$, where $\zeta$ and $n$ arise in the Taylor expansion of $\sqrt{\Delta}(r,0)$ around $r=0$ as
\begin{equation}
    \sqrt{\Delta}(r,0)=\zeta r^{n-1}+.....,
\end{equation}
and $2\alpha$ is the coefficient of $r^3$ in the expression of the Misner-Sharp mass function $F$ in the $(r,R)$ coordinates. One can choose a suitable $\mathcal{Y}$ and $F$, due to two degrees of freedom available to us, such that the end state of a non-zero pressured collapsing cloud, satisfying the weak energy condition and the regularity condition, is globally visible. Let us consider the Misner-Sharp mass function of the form
\begin{equation}
F(r,R)=F_0\left(\frac{r}{r_c}\right)^3+F_3\left(\frac{r}{r_c}\right)^6+F_R\left(\frac{R}{r_c}\right)^3.
\end{equation}
We set $\mathcal{Y}=1$, which corresponds to an acceleration-free marginally bound case. The total mass of the collapsing cloud is $(F_0+F_3+F_R)/2$.

It should be noted that for the collapsing fluid to satisfy the weak energy condition, one of the necessary criteria is to have the mass function to be a monotonically increasing function of the areal coordinate $R$ in the $(r,R)$ frame. Hence, one realizes that the formalism discussed in 
\cite{Giambo, Giambo2} 
is restricted to studying the end state of only those collapsing fluid with negative radial pressure, as seen in Eq.(\ref{prgiambo}). It can be seen that for suitable values taken by the components of the Misner-Sharp mass function, one gets a globally visible end-state singularity. 
\section{Concluding Remarks}
Some concluding points and concerns are discussed below:
\begin{enumerate}

\item A non-zero measured set of initial parameters, which relate to the total mass and the initial mean density of the collapsing cloud, is obtained, which leads to a globally visible singularity as its end state. The Misner-Sharp mass function is chosen such that the singularity formed is physically strong in the sense of Tipler. Hence, it can be concluded that such globally visible singularity is stable under small perturbations in the initial data ($\frac{F_0}{4}$ and $\Bar{\rho_0}$) of the collapsing spherical cloud. 

\item Globally visible singularity is possible to achieve in principle; however, here, we have discussed the possibility of its existence in an astrophysical scenario by looking into the possibility of occurrence of suitable mass and size required for its existence.  The singularity at the center of the M87 galaxy may be globally visible if the initial radius of the collapsing cloud forming it is more than $27.05$ times its Schwarzschild radius. Similarly, the collapsing cloud's initial radius forming the SgrA* singularity should be more than $26.42$ times its Schwarzschild radius to end up being globally visible. Here, we have modeled the collapsing cloud to be marginally bound, spherically symmetric, and pressureless. The outcome of the causal structure of the singularity may vary if these assumptions are dropped.

\item If a density contrast $\frac{\delta \rho}{\rho}$ of order one is achieved in the matter-dominated era just after the time of matter-radiation equality, the configuration detaches from the background universe and starts contracting due to gravity. A configuration having a total mass of the order less than $10^{45}$ kg forms a physically strong (in the sense of Tipler) singularity, which can be visible globally. It should be noted that we have not taken into account the process of virialization, which is an essential phenomenon in cosmology, and which could oppose the unhindered gravitational collapse of the configuration during the primordial time, causing it to attain an equilibrium, thereby avoiding the formation of the singularity at all.

\item A neutron star, which is a part of the binary system, can reach a critical mass by accreting the supernova ejecta of the companion exploding star, after which it can collapse unhindered. We consider four such models having different critical masses and their corresponding radii. Each of these models is obtained by fixing six nuclear parameters, as discussed in 
\cite{Belvedere}. 
The final state of such accreting neutron star is found to be a singularity that is globally hidden for all these models. One may claim that these act as positive evidence for the hypothesis of weak cosmic censorship. However, a loophole in this argument, which is of concern, is that the collapse formalism used for our purpose assumes zero pressure; however, neutron stars may have non-zero pressure satisfying the polytropic equation of state. To address this problem, we have incorporated the formalism proposed by Giambo \textit{et. al.} 
\cite{Giambo, Giambo2}. 
However, the formalism works only for negative radial pressure if one has to satisfy the weak energy condition. One could obtain a suitable mass function $F(r,R)$ and the function $\mathcal{Y}(r,R)$ for the collapse of supernova ejecta accreting neutron star after achieving the critical mass and obtain a globally naked singularity as the end state of the collapse, as seen in Fig.(4). This end state is stable under small perturbation in the initial data (here $F_0$ and $F_3$).  The formalism developed in 
\cite{Giambo, Giambo2} 
works only for the subset of configurations that have negative radial pressure. The formalism of finding out the global causal structure of the singularity formed due to gravitational collapse is not developed enough to incorporate arbitrary pressures and equation of state as demanded by the neutron star's complicated structure. A simplified model giving rise to both the possibilities: hidden and visible, globally, indicates that even a more realistic model may give rise to a globally visible singularity.

\item An external observer encountering an escaping singular null geodesic may find traces of quantum gravity encoded in it, making such singularities more tempting to investigate. We have shown here that the formation of a globally visible singularity may also arise in the astrophysical setup and is not merely a mathematical artifact. One could therefore interpret that such singularities, after all, may not be so elusive.  

\end{enumerate}

\section{Acknowledgement}

KM would like to acknowledge the support of
the Council of Scientific and Industrial Research (CSIR, India, Ref: 09/919(0031)/2017-EMR-1) for funding the work.

\section{Appendix}
Here, we discuss the local visibility of the singularity formed due to a marginally as well as non- marginally bound collapse of the dust sphere, The Taylor expansion of the time curve around $r=0$ is expressed as
\begin{equation}
     t(r,v)=t(0,v)+r\chi_1(v)+r^2\chi_2(v)+r^3\chi_3(v)+O(r^4),
\end{equation}
where
\begin{equation}\label{chii}
    \chi_i(v)=\frac{1}{i!}\frac{d^i t}{dr^i}\bigg |_{r=0}.
\end{equation}
Here $v$ is called the scaling function and is related to the physical radius and the comoving radius as 
\begin{equation}
    R(t,r)=r v(t,r).
\end{equation}
Coordinate freedom gives us the option of rescaling the physical radius as
\begin{equation}
    R(t_i,r)=r,
\end{equation}
where $t_i$ is the initial time of the collapse. This is equivalent to having $v(t_i,r)=1$. First three $\chi_i$'s can be calculated by integrating Eq.(\ref{friedmann}), and is expressed as 
\cite{Joshi3, Mosani2}:
\begin{equation}\label{chi1}
    \chi_1(v)=-\frac{1}{2}\int^{1}_{v} \frac{\frac{G \mathcal {F}_1}{ v}+f_1c^2}{\left(\frac{G \mathcal{F}_0}{ v}+f_0c^2\right)^{\frac{3}{2}}}dv, 
\end{equation}
\begin{equation}\label{chi2}
    \chi_2(v)=\int^{1}_{v}\left[\frac{3}{8}\frac{\left(G\frac{\mathcal{F}_1}{v}+f_1 c^2\right)^2}{\left(G\frac{\mathcal{F}_0}{v}+f_{0}c^2\right)^{\frac{5}{2}}}-\frac{1}{2}\frac{\frac{G\mathcal{F}_2}{v}+f_{2}c^2}{\left(\frac{G\mathcal{F}_0}{v}+f_{0}c^2\right)^{\frac{3}{2}}}\right] dv
\end{equation}
and
\begin{widetext}
\begin{equation}\label{chi3}
     \chi_3(v)=  \int_v^1 \frac{f_{1}}{\left(\frac{G \mathcal{F}_0}{v}+f_{0}\right)^{\frac{3}{2}}}\left(-\frac{5}{16}\left(\frac{f_{1}}{\frac{G\mathcal{F}_0}{v}+f_{0}c^2}\right)^2 + \frac{3}{4}\left(\frac{\frac{G \mathcal{F}_2}{v}+f_{2}c^2}{\frac{G\mathcal{F}_0}{v}+f_{0}c^2}\right)\right)-\frac{1}{2}\frac{\left(\frac{G\mathcal{F}_3}{v}+f_{3}c^2\right)}{\left(G\frac{\mathcal{F}_0}{v}+f_{0}c^2\right)^{\frac{3}{2}}}dv.
\end{equation}
\end{widetext}
Here, $f_i$ and $\mathcal{F}_i$ are such that
\begin{equation}
F= \sum_{i=0}^{\infty} \mathcal{F}_i r^{i+3} 
\end{equation}
and
\begin{equation}
    f=\sum_{j=0}^{\infty}f_j r^{j+2}.
\end{equation}
near $r=0$. It was shown in 
\cite{Mosani} 
that for a strong singularity to be locally naked, we should have $\chi_1=\chi_2=0$ and $\chi_3>0$ at $v=0$. This condition is always satisfied for the case of marginally bound collapse with the collapsing fluid having mass function as in Eq.(\ref{strongmassfunction}) with $F_3<0$, as is apparent from Eq.(\ref{chi1}-\ref{chi3}). The condition may or may not satisfy for a non zero velocity function. However, for the velocity function $f=10^{-5}\frac{G F}{c^2r}$, as chosen in Fig.(1b), the above mentioned condition is satisfied. Hence, all the singularities arising in this article are at least locally naked. 
 

\begin{thebibliography}{}
 
 \bibitem{Akiyama} K. Akiyama \textit{et al.} [Event Horizon Telescope Collabora-
tion], \href{https://iopscience.iop.org/article/10.3847/2041-8213/ab0ec7}{Astrophys. J. \textbf{875}, 1 (2019)}.

\bibitem{Schodel} R. Schodel \textit{et. al.}, \href{https://www.nature.com/articles/nature01121}{Nature, \textbf{419}, 694 (2002)}.

\bibitem{Ghez} A. M. Ghez, \textit{et. al.}, \href{https://ui.adsabs.harvard.edu/abs/2008ApJ...689.1044G/abstract}{Astrophys. J., \textbf{689}, 1044 (2008)}.

\bibitem{Kormendy} J. Kormendy and L. C. Ho, \href{https://www.annualreviews.org/doi/10.1146/annurev-astro-082708-101811}{ARAA \textbf{51}, 511 (2013)}.

\bibitem{Hawking} S. W. Hawking and G. F. R. Ellis, The large scale structure of space-time, Cambridge University Press (1973).

\bibitem{Penrose} R. Penrose, Riv. Nuovo Cimento Soc. Ital. Fis. \textbf{1}, 252 (1969).

S.W. Hawking and G. F. R. Ellis, The Large Scale Structure
of Spacetime (Cambridge University Press, Cambridge,
United Kingdom, 1973).

\bibitem{Deshingkar} S. S. Deshingkar, S. Jhingan and P. S. Joshi, \href{https://link.springer.com/article/10.1023%2FA%3A1018813108516
}{Gen. Relativ. Gravit. \textbf{30}, 1477 (1998)}.

\bibitem{Mosani} K. Mosani, D. Dey and P. S. Joshi, \href{https://arxiv.org/abs/2003.07092}{arXiv:2003.07092v1}

\bibitem{Shaikh} R. Shaikh, P. Kocherlakota, R. Narayan, and P. S. Joshi, \href{https://academic.oup.com/mnras/article-abstract/482/1/52/5113467?redirectedFrom=fulltext}{Mon. Not. Roy. Astron. Soc. \textbf{482}, 52 (2019)}.

\bibitem{JMN} P. S. Joshi, D. Malafarina and R. Narayan, \href{https://iopscience.iop.org/article/10.1088/0264-9381/28/23/235018}{Classical Quantum Gravity, \textbf{28}, 23 (2011)}.

\bibitem{JNW} A. I. Janis, E. T. Newman, and J. Winicour \href{https://journals.aps.org/prl/abstract/10.1103/PhysRevLett.20.878}{
Phys. Rev. Lett. \textbf{20}, 878 (1968)}. 

\bibitem{Bambhaniya} P. Bambhaniya, A. B. Joshi, D. Dey, and P. S. Joshi, \href{https://journals.aps.org/prd/abstract/10.1103/PhysRevD.100.124020}{Phys. Rev. D \textbf{100}, 124020 (2019)}.

\bibitem{AJoshi} A. B. Joshi, P. Bambhaniya, D. Dey, and P. S. Joshi, \href{https://arxiv.org/abs/1909.08873}{arXiv: 1909.08873}.

\bibitem{Dey} D. Dey, P. S. Joshi, A. Joshi, and P. Bambhaniya, \href{https://www.worldscientific.com/doi/abs/10.1142/S0218271819300246}{Int. J.
Mod. Phys. D \textbf{28}, no. 14, 1930024 (2019)}.

\bibitem{Dey2} D. Dey, K. Bhattacharya, and T. Sarkar, \href{ https://journals.aps.org/prd/abstract/10.1103/PhysRevD.87.103505}{Phys. Rev. D \textbf{87}, 103505 (2013).}

\bibitem{Dey3} D. Dey, K. Bhattacharya, and T. Sarkar, \href{https://journals.aps.org/prd/abstract/10.1103/PhysRevD.88.083532}{
Phys. Rev. D \textbf{88}, 083532 (2013)}.


\bibitem{Shaikh2} R. Shaikh and P. S. Joshi, \href{https://iopscience.iop.org/article/10.1088/1475-7516/2019/10/064/meta}{Journal of Cosmology and Astroparticle Physics \textbf{10}, 064 (2019)}.

\bibitem{Bambi} A. B. Abdikamalov, A. A. Abdujabbarov, D. Ayzenberg, D. Malafarina, C. Bambi, and B. Ahmedov, \href{https://journals.aps.org/prd/abstract/10.1103/PhysRevD.100.024014}{
Phys. Rev. D \textbf{100}, 024014 (2019)}.

\bibitem{Dey4} K. Bhattacharya, D. Dey, A. Mazumdar, and T. Sarkar, \href{ https://journals.aps.org/prd/abstract/10.1103/PhysRevD.101.043005}{Phys. Rev. D \textbf{101}, 043005 (2020)}.

\bibitem{Wheeler} J. A. Wheeler, in Relativity, groups and topology, Les Houches, eds. C. DeWitt, and B. DeWitt, Gordon and Breach, New York (1964).

\bibitem{HawkingCUP} S. W. Hawking, In General Relativity: An Einstein Centenary Survey, eds S. W. Hawking and W. Israel. Cambridge: Cambridge University Press (1979).

\bibitem{Bergmann} P. G. Bergmann, Some strangeness in the proportion, eds. H. Woolf, Addison-Wesley Reading, Massachusetts (1980).

\bibitem{JoshiCUP} P. S. Joshi, Gravitational Collapse and Spacetime Singularities (Cambridge University Press, Cambridge, England, 2007).

\bibitem{Lemaitre} G Lemaitre, \href{https://ui.adsabs.harvard.edu/abs/1933ASSB...53...51L/abstract}{Ann. Soc. Sci. Bruxelles I A \textbf{53}, 51 (1933)}.

\bibitem{Tolman} R. C. Tolman, \href{https://www.pnas.org/content/20/3/169}{Proc. Natl. Acad. Sci. USA \textbf{20}, 410 (1934)}. 

\bibitem{Bondi} H. Bondi, \href{https://academic.oup.com/mnras/article/107/5-6/410/2601230}{Mon. Not. Astron. Soc. \textbf{107}, 343 (1947)}.

\bibitem{Misner} C. W. Misner and D. H. Sharp, \href{https://journals.aps.org/pr/abstract/10.1103/PhysRev.136.B571}{Phys. Rev. \textbf{136}, B571 (1964)}. 

\bibitem{Ellis} G.F.R. Ellis and B. G. Schmidt, \href{https://link.springer.com/article/10.1007/BF00759240}{ Gen. Relativ. Gravit.\textbf{8}, 915 (1977)}.

\bibitem{Tipler} F. J. Tipler, \href{https://www.sciencedirect.com/science/article/abs/pii/0375960177905084}{Phys. Lett. \textbf{64A}, 8 (1977)}.

\bibitem{Clarke} C. J. S. Clarke and A. Krolak, \href{https://www.sciencedirect.com/science/article/pii/0393044085900129}{J. Geo. Phys. \textbf{2}, 127 (1986)}.

\bibitem{Joshi} P. S. Joshi, I. H. Dwivedi,\href{https://journals.aps.org/prd/abstract/10.1103/PhysRevD.47.5357}{Phys. Rev. D \textbf{47}, 5357 (1993)}.

\bibitem{Joshi2} P. S. Joshi, Global Aspects in Gravitation and Cosmology (Clendron Press, Oxford, 1993).

\bibitem{Joshi3} P. S. Joshi, D. Malafarina, and R. V. Saraykar, \href{https://www.worldscientific.com/doi/abs/10.1142/S0218271812500666}{Int. J. Mod. Phys. D. \textbf{21}, 08, 1250066 (2012)}.

\bibitem{Mosani2} K. Mosani, D. Dey, and P. S. Joshi, \href{https://journals.aps.org/prd/abstract/10.1103/PhysRevD.101.044052}{Phys. Rev. D \textbf{101}, 044052 (2020)}.


\bibitem{Toropina} O. D. Toropina, M. M. Romanova, and R. V. E. Lovelace, \href{https://academic.oup.com/mnras/article/420/1/810/1050080}{Mon. Not. Roy. Astron. Soc., \textbf{420}, 810 (2012).}

\bibitem{Hoyle} H. Bondi and F. Hoyle, \href{https://ui.adsabs.harvard.edu/abs/1944MNRAS.104..273B/abstract}{Mon. Not. Roy. Astron. Soc., \textbf{104}, 273 (1944).}

\bibitem{Chevalier} R. A. Chevalier, \href{https://ui.adsabs.harvard.edu/abs/1989ApJ...346..847C/abstract}{Astrophys. J \textbf{346}, 847 (1989).}


\bibitem{Belvedere} R. Belvedere, D. Pugliese, J. A. Rueda, R. Ruffini, and S.-S. Xue, \href{https://www.sciencedirect.com/science/article/abs/pii/S0375947412001029}{Nuclear Phys. A., \textbf{883}, 1 (2012).}




\bibitem{Gillessen} S.Gillessen, F. Eisenhauer, S. Trippe, T. Alexander, R. Genzel, F. Martins, and T. Ott, \href{https://iopscience.iop.org/article/10.1088/0004-637X/692/2/1075}{The Astrophysical Journal, \textbf{692}, 2 (2009)}.

\bibitem{Akiyama2} K. Akiyama \textit{et al.} [Event Horizon Telescope Collabora-
tion], \href{https://iopscience.iop.org/article/10.3847/2041-8213/ab1141}{The Astrophysical Journal Letters, \textbf{875}, 1 (2019)}.

\bibitem{Zeldovich} Y. B. Zel’dovich and I. D. Novikov, \href{https://ui.adsabs.harvard.edu/abs/1967SvA....10..602Z/abstract}{Sov. Astron. \textbf{10}, 602 (1967)}.

\bibitem{Hawking2} S. Hawking, \href{https://academic.oup.com/mnras/article/152/1/75/2604549}{Mon. Not. R. Astron. Soc. \textbf{152}, 75 (1971)}.

\bibitem{Frieman} J. A. Frieman, M. S. Turner, and D, Huterer, \href{https://www.annualreviews.org/doi/10.1146/annurev.astro.46.060407.145243}{Annual Review of Astronomy and Astrophysics \textbf{46}, 385 (2008)}.

\bibitem{Farooq} O. Farooq, F. Ranjeet Madiyar, S. Crandall, and B. Ratra, \href{https://iopscience.iop.org/article/10.3847/1538-4357/835/1/26}{The Astrophysical Journal, \textbf{835}, 1 (2016)}.

\bibitem{Liddle} A. Liddle, An introduction to modern cosmology, Wiley publication (2003).

\bibitem{Planck} Planck Collaboration, \href{https://arxiv.org/abs/1807.06209}{arXiv:1807.06209}.


\bibitem{Khlopov} M. Y. Khlopov and A. G. Polnarev, \href{https://linkinghub.elsevier.com/retrieve/pii/0370269380906243}{Phys. Lett. B, \textbf{97},3-4 (1980)}.

\bibitem{Reuda} J. A. Rueda and R. Ruffini, \href{https://iopscience.iop.org/article/10.1088/2041-8205/758/1/L7}{Astrophys. J.,  \textbf{758}, 1 (2012).}

\bibitem{Lalazissis} G. A. Lalazissis, J. König, and P. Ring, \href{https://journals.aps.org/prc/abstract/10.1103/PhysRevC.55.540}{
Phys. Rev. C \textbf{55}, 540 (1997).}

\bibitem{Sharma} M.M.Sharma, M.A.Nagarajan, and P.Ring, \href{https://www.sciencedirect.com/science/article/abs/pii/037026939390970S}{Phys. Lett. B, \textbf{312}, 377 (1993).}

\bibitem{Sugahara} Y. Sugahara and H. Toki, \href{https://ui.adsabs.harvard.edu/abs/1994NuPhA.579..557S/abstract}{Nuclear Phys. A \textbf{579}, 557 (1994).}

\bibitem{Hirata} D. Hirata, H. Toki, and I. Tanihata, \href{https://www.sciencedirect.com/science/article/abs/pii/037594749500035Y}{Nuclear Phys. A, \textbf{589}, 239 (1995).}

\bibitem{Giambo} R. Giambò, F. Giannoni, G. Magli and P. Piccione, \href{https://link.springer.com/article/10.1007/s00220-003-0793-9}{Communications in Mathematical Physics \textbf{235}, 545–563 (2003).}

\bibitem{Giambo2} R. Giambò, \href{https://aip.scitation.org/doi/10.1063/1.2167919}{Journal of Mathematical Physics \textbf{47}, 022501 (2006).} 
\end{thebibliography}
\end{document}